\documentclass[aps,prl,twocolumn,showpacs,preprintnumbers,amsmath,superscriptaddress,amssymb,floats,nofootinbib]{revtex4}

\usepackage{graphicx}
\usepackage{dcolumn}

\begin{document}

\title{Search for Spin-Dependent Short-Range Force Using Optically Polarized $^3$He Gas}

\author{W. Zheng}
\author{H. Gao} 
\author{B. Lalremruata}
\author{Y. Zhang}
\author{G. Laskaris}
\affiliation{Triangle Universities Nuclear Laboratory and Department of Physics, Duke University, Durham, North Carolina 27708, USA}
\author{W.M. Snow}
\affiliation{Indiana University, Bloomington, Indiana 47408, USA}
\author{C.B. Fu}
\affiliation{Department of Physics, Shanghai Jiaotong University, Shanghai, 200240, China}
 
\begin{abstract}
We propose a new method to detect short-range \textit{P-} and \textit{T-} violating interactions between nucleons, based on measuring the precession frequency shift of polarized $^3$He nuclei in the presence of an unpolarized mass. To maximize the sensitivity, a high-pressure $^3$He cell with thin glass windows (250 $\rm\mu m$) is used to minimize the distance between the mass and $^3$He. The magnetic field fluctuation is suppressed by using the $^3$He gas in a different region of the cell as a magnetometer. Systematic uncertainties from the magnetic properties of the mass are suppressed by flipping both the magnetic field and spin directions. Without any magnetic shielding, our result has already reached the sensitivity of the current best limit. With improvement in uniformity and stability of the field, we can further improve the sensitivity by two orders of magnitude over the force range from $10^{-4}-10^{-2}$ m.
 
\end{abstract}

\pacs{13.88.+e, 13.75.Cs, 14.20.Dh, 14.70.Pw}

\maketitle

The possible existence of new forces with weak couplings and macroscopic ranges have been proposed by several authors \cite{Leitner,Hill,Dobrescu,Jaeckel}. A \textit{P}- and \textit{T}- violating macroscopic force with an interaction range from cm to $\rm\mu m$ first proposed in \cite{Moody} has the form
\begin{equation}
	V(z)=\frac{g_sg_p\hbar^2\hat{\sigma}\cdot\hat{r}}{8\pi m_n}(\frac{1}{r\lambda}+\frac{1}{r^2})\exp(-r/\lambda),
	\label{eq:potential}
\end{equation}
where $g_s$ and $g_p$ are the scalar and pseudoscalar coupling constants, $\hbar$ is the Plank's constant, $\hat{\sigma}$ is the spin of the polarized nucleon, $\hat{r}=\vec{r}/r$ is the unit vector from the unpolarized nucleon to the polarized nucleon, $m_{n}$ is the nucleon mass, and $\lambda$ is the range of the force. This short-range force is mediated by exchanging an axion-like particle between unpolarized nucleons and polarized nucleons. A similar interaction may also exist between nucleon and electron. 
Many experimental efforts have been devoted to search for this interaction between either nucleons or nucleon and electron, and various techniques have been used, such as sensitive torsion pendula \cite{Ritter,Hammond}, clock comparisons between two different polarized species \cite{Youdin,Vasilakis,Glenday}, and measurements of neutron bound states on a flat surface in the gravitational field \cite{Baessler}. Very recently, measurements of the longitudinal relaxation rate $\Gamma_1$ and transverse relaxation rate $\Gamma_2$ of polarized $^3$He gas were used to search for this short-range interaction between nucleons \cite{Pokotilovski,Serebrov,Fu,Petukhov}. As the relaxation time of polarized $^3$He can be as long as tens of hours, any new interaction with the polarized $^{3}$He nuclei can lead to a visible change in the relaxation time. These measurements provide to our knowledge the most stringent direct laboratory constraint on the coupling constant product $g_sg_p$ for a monopole-dipole interaction between nucleons of the form in Eq. (\ref{eq:potential}) over distances from $10^{-6}$ to $10^{-2}$ m \cite{Petukhov}. Note that this limit is still more than 9 orders of magnitude larger than the standard Axion coupling originally proposed to solve the strong CP problem \cite{Youdin,Peccei}. 

In this work, we present a new method to search for the spin-dependent macroscopic force between nucleons by measuring the frequency difference of optically polarized $^3$He gas with and without a nearby unpolarized mass. The frequency difference due to the magnetic field gradient is a first order effect, as such it is more sensitive than the relaxation measurement because the gradient-induced relaxation is a second order effect \cite{Slichter,Cates}. We also performed a pilot experiment to demonstrate how this method works. With a modest stability of the magnetic field, the sensitivity of this experiment already reaches the current best laboratory limit on $g_sg_p$. With improved stability of the magnetic field, the proposed method could be used to improve the current best limit by two to three orders of magnitude in the force range from $10^{-4}$ to $10^{-2}$ m. 

The spin-dependent short-range interaction changes the precession frequency of the polarized nuclei through the interaction $\hat{\sigma}\cdot\hat{r}$ in Eq. (\ref{eq:potential}), which is similar to the well-known $\hat{\mu}\cdot\vec{B}$ interaction of a magnetic dipole moment $\hat{\mu}$ in an external magnetic field $\vec{B}$. Consider a cylindrical cell containing polarized $^3$He gas with its polarization pointing in the $z$ direction along the axis of the cylinder and a block of unpolarized mass is placed next to the end of the cell. The short-range interaction on each $^3$He atom inside the cell can be obtained by integrating Eq. (\ref{eq:potential}) over the unpolarized source mass. In the limit case in which the transverse dimensions ($x,y$ direction) of the mass are much larger than the force range, the mass can be approximated as an infinite plane source with its normal pointing in the $z$ direction. In this limit the frequency shift from the planar mass block is \cite{Fu}
\begin{equation}
	\Delta\omega=\frac{g_sg_pN\hbar\lambda}{4m_n}e^{-z/\lambda}(1-e^{-d_0/\lambda}),
	\label{eq:omega}
\end{equation}
where $z$ is the distance measured from the surface of the mass block to the polarized $^3$He, $N$ is the nucleon number density of the mass, and $d_0$ is the thickness of the mass block. If the external magnetic field is uniform with a value of $B_0$, then the spin precession frequency in the presence of the mass has a spatial dependence which can be written as 
\begin{equation}
	\omega=\omega_0+Ae^{-z/\lambda},
\end{equation}
where $A=\frac{g_sg_pN\hbar\lambda}{4m_n}(1-e^{-d_0/\lambda})$. As the precession signal received by the pickup coil is a weighted sum from all the $^3$He inside the cell, the signal induced in the pickup coil is
\begin{equation}
	S\propto\int^{\infty}_{d}\cos(\omega_0t+Ae^{-z/\lambda}t)B(z)dz,
	\label{eq:signal}
\end{equation}
where $d$ is the window thickness of the cell, $B(z)$ is the field profile of the pickup coil along the cell axis, and the reciprocity theorem is applied here to compute the signal induced in the pickup coil \cite{Insko}. When $A$ is zero, Eq. (\ref{eq:signal}) is a pure sinusoidal function with a well-defined frequency. When $A$ is nonzero Eq. (\ref{eq:signal}) shifts the mean frequency of the signal. The mean oscillation frequency determined from $N_c$ observed periods during a time $T$ is $f=N_c/T$. In the presence of the interaction for the same number of periods the time duration changes to $T^{\prime}$ and the new frequency is $f^{\prime}=N_c/T^{\prime}$. The frequency difference is~\footnote{When S/N becomes the limiting factor of the measurement (not the case in the present work), phase difference or frequency spectrum may provide a better measurement of the frequency shift than the peak counting method.}
\begin{equation}
	\Delta f=\frac{N_c}{T^{\prime}}-\frac{N_c}{T}\cong-f\frac{\Delta T}{T},
	\label{eq:dfdT}
\end{equation}
where $\Delta T=T^{\prime}-T$. Eq. (\ref{eq:dfdT}) establishes a relation between $\Delta f$ and $\Delta T$. $\Delta T$ can be calculated for different strengths $g_{s}g_{p}$ and ranges $\lambda$ of the spin-dependent interaction by numerically integrating Eq. (\ref{eq:signal}). For any given $\lambda$, the parameter $A$ in Eq. (\ref{eq:signal}) is tuned in such a way that the calculated $\Delta T$ matches the experimentally determined frequency shift $\Delta f$. Hence, constraints on $g_sg_p$ with different values of $\lambda$ are established given the sensitivity of the experiment.

\begin{figure}[htbp]
	\centering
		\includegraphics[width=0.4\textwidth]{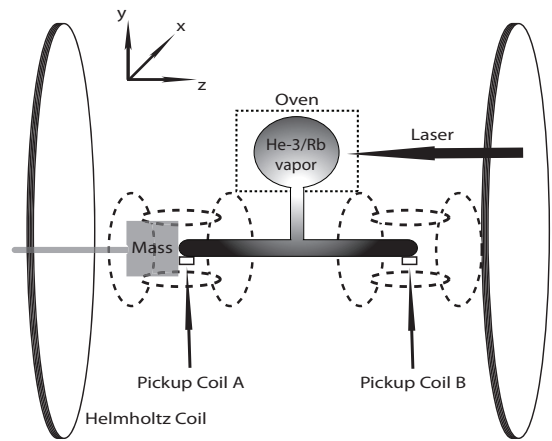}
	\caption{The diagram of the test experiment apparatus (not to scale). The cylindrical cell axis is the $z$ direction. The cell contains 7 amg $^3$He gas and is optically pumped in the pumping chamber to about 40$\%$ polarization. The coils in the dashed lines are gradient coils to actively compensate the gradients from the Helmholtz coil and other background fields.}
	\label{fig:Expsetup}
\end{figure}

In this pilot experiment we used a 7 amg \footnote{1 amg is the number density of 1 atm gas molecules at 0 $^{\rm o}$C.} high pressure $^3$He cell originally constructed as a $^3$He gas target for two- and three-body photo-disintegration experiments \cite{Ye}. The cell had two chambers, a spherically-shaped spin-exchange optical pumping chamber and a 40 cm long cylindrical target chamber connected by a glass tube (see Fig. \ref{fig:Expsetup}). The target chamber had two thin glass windows on its ends. The thickness of the window was about 250 $\rm\mu m$. A Macor machinable ceramic mass block was used as the unpolarized mass. It was repeatedly brought into contact with and moved away from the cell window by a G-10 rod connected to a stepper motor. The stepper motor moved the mass to a final position with better than $10$ micron repeatability, more than an order of magnitude smaller than the cell window thickness. Two identical pickup coils were mounted right below each window. Pickup coil A was used to measure the frequency of the polarized $^{3}$He nuclei influenced by the mass; pickup coil B was used to monitor the magnetic holding field. The 40 cm long $^3$He cell was positioned at the center of a Helmholtz coil pair. Due to fairly large gradients at the end of the cylindrical chamber, two identical two-axis gradient coil pairs (dashed line in Fig. \ref{fig:Expsetup}) were added at both ends to compensate the gradients from the Helmholtz coils and background fields. The measured transverse relaxation time $T_2$ of the $^3$He signal is greater than 1 s with this arrangement.  
   
Free induction decay (FID) at 24 kHz was performed to measure the $^3$He precession signal. The RF pulse with small tipping angle was applied to make the polarization loss negligible. The precession signal is digitized and recorded by the computer. In order to determine the frequency unambiguously, the acquisition time stopped at the instant when the signal-to-noise ratio is either below 10 or at 0.2 s, whichever comes first. The frequency is computed by counting the periods during the acquisition time. In a 7 amg $^3$He gas cell, the diffusion constant is about 0.27 cm$^2$/s \cite{Saam}, and it is known that the effective diffusion rate is lowered at the cell boundary \cite{Swiet}. Therefore effects from the diffusion of the $^3$He can be ignored during the FID measurement and it is valid to use Eq. (\ref{eq:signal}) to analyze the experimental data. Each measurement cycle contains two measurements: first with the mass in contact with the window (in position) and next with the mass moved far away from the window (out position). Simultaneous measurements at pickup coil B were performed to monitor the magnetic field fluctuations. The peak to peak variation of the field is about $5\times10^{-3}\%$. Removing the field fluctuations measured by coil B reduced the peak to peak variation of the field to $4\times10^{-4}\%$. After the field correction, the frequency difference between the ``in'' and ``out'' positions is calculated as $\Delta f=f_{in}-f_{out}$. 

The magnetic susceptibility of the mass can change the field at the location of the $^{3}$He through its effect on the holding field and therefore lead to a systematic effect. Although the magnetic susceptibility of Macor ceramic is known to be small enough not to cause a systematic error in this test experiment, the real material can in principle contain paramagnetic or even ferromagnetic impurities. Paramagnetic impurities would increase the local field strength (and therefore the $^{3}$He precession frequency) independent of field direction. The spin-dependent interaction can increase or decrease the precession frequency depending on the magnetic holding field direction. We therefore can isolate a possible spin-dependent interaction from paramagnetic effects by flipping the magnetic field. However, a frozen-in field from possible ferromagnetic impurities has the same magnetic field dependence as the spin-dependent interaction, which makes it difficult to separate them apart. The most likely ferromagnetic contamination of the mass block comes from machining process, during which ferromagnetic tools are usually used to cut the material. To minimize this effect, cutting tools with diamond tips are used to ensure that there is no physical contact between the metallic part of the tools and the surface of the block. Additionally, we also flip spin direction in order to cancel any spin-dependent systematic effect in the system if there is any. Therefore, we took data in four different configurations for B field and spin directions, $\Delta f_{++}$, $\Delta f_{-+}$, $\Delta f_{--}$, and $\Delta f_{+-}$, representing B/S: ++, B/S: -+, B/S: --, and B/S: +-, respectively. If $\Delta f_B$ represents the field-dependent frequency shift, $\Delta f_S$ represents spin-dependent frequency shift, and $\Delta f_0$ represents frequency shift without B or S dependence, then
\begin{align}
	\Delta f_{++} &= +\Delta f_B + \Delta f_S + \Delta f_0 \\
	\Delta f_{-+} &= -\Delta f_B + \Delta f_S + \Delta f_0 \\
	\Delta f_{--} &= -\Delta f_B - \Delta f_S + \Delta f_0 \\
	\Delta f_{+-} &= +\Delta f_B - \Delta f_S + \Delta f_0. 
\end{align}
In this notation, the short-range force induced $\Delta f_B$ can be expressed as
\begin{equation}
	\Delta f_B =\frac{1}{4}(\Delta f_{++}+\Delta f_{+-}-\Delta f_{--}-\Delta f_{-+}).
	\label{eq:fb}
\end{equation}
The uncertainty of $\Delta f_B$ is given by
\begin{equation}
	\frac{1}{4}\sqrt{\sigma_{++}^2+\sigma_{+-}^2+\sigma_{--}^2+\sigma_{-+}^2},
\end{equation}
where $\sigma_{++}$, $\sigma_{-+}$, $\sigma_{--}$, and $\sigma_{+-}$ are the uncertainties of $\Delta f_{++}$, $\Delta f_{-+}$, $\Delta f_{--}$, and $\Delta f_{+-}$, respectively. The noise in the measurement is mainly due to the magnetic field fluctuation, which limits the uncertainties of the frequency measurements.

\begin{figure}[htbp]
	\centering
		\includegraphics[width=0.46\textwidth]{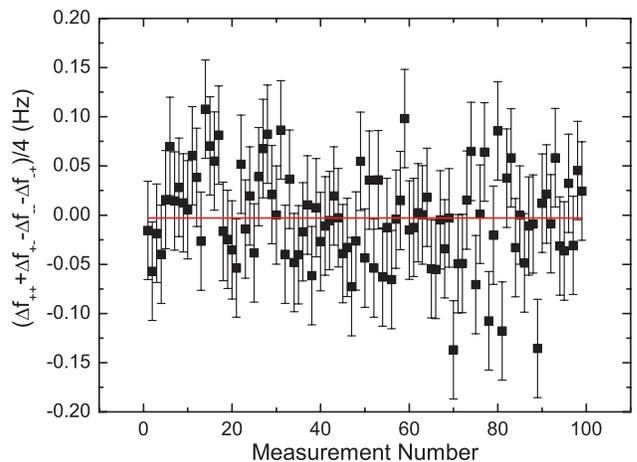}
		\caption{(Color online) The frequency difference correlated with the position of the ceramic mass block. The error bars show the standard deviation of the magnetic holding field after correction by coil B.}
	\label{fig:ceramicresult}
\end{figure}

We performed $100$ measurement cycles for each of the four configurations to determine the average frequency difference between the two mass positions. The frequency differences from the spin-dependent force for these runs are shown in Fig. \ref{fig:ceramicresult}. The average frequency difference of 100 measurements was $\Delta f_B=-0.003\pm0.005$ Hz, consistent with zero. 

\begin{figure}
	\centering
		\includegraphics[width=0.45\textwidth]{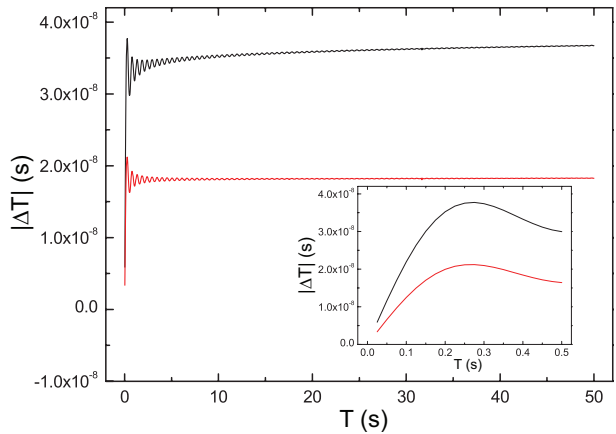}
	\caption{(Color online) The upper curve shows the $\left|\Delta T\right|$ as a function of $T$, using the real field profile of the pickup coil. The lower curve shows the same curve but with the pickup coil profile approximated by a rectangular function. The inset of the figure shows the linear behavior of $\left|\Delta T\right|$ at small $T$.}
	\label{fig:dt-T}
\end{figure}

Eq. (\ref{eq:fb}) is then used to calculate the frequency difference from the spin-dependent short-range force to place an upper limit on $g_sg_p$ as a function of $\lambda$. For $f=24$ kHz and $\Delta f=0.005$ Hz, Eq. (\ref{eq:dfdT}) yields $\Delta T=-2.1\times 10^{-7}T$ s. With a fixed frequency shift, $\Delta T$ should increase linearly with respect to the acquisition time $T$. Theoretically, $\Delta T$ is obtained by comparing Eq. (\ref{eq:signal}) with a sinusoidal function in absence of the force. With a real pick-coil profile, Eq. (\ref{eq:signal}) is obtained by numerical integration, using the actual geometry of the experiment. The resultant $\Delta T$ as a function of $T$ is shown in Fig. \ref{fig:dt-T} (the upper black curve). Surprisingly, $\Delta T$ increases linearly only for a short period of time. As time elapses, $\Delta T$ oscillates around a constant value, indicating that the frequency shift due to this exponential type of force is not fixed and diminishes at large $T$. This striking behavior suggests that one will not gain more information from longer measurement time though the frequency resolution is improved by doing so.  

An closed-form solution of Eq. (\ref{eq:signal}) can be obtained if one approximates the real profile of $B(z)$ by a rectangular function with a cut-off position $w$ mimicing the width of the profile. In this case, the upper limit of the integral is replaced by $w$, and the integration yields
\begin{align}
	S(t)=\lambda[\cos(\omega_0t)(\textrm{Ci}(Ae^{-\frac{d}{\lambda}}t)-\textrm{Ci}(Ae^{-\frac{w}{\lambda}}t))\notag \\
	-\sin(\omega_0t)(\textrm{Si}(Ae^{-\frac{d}{\lambda}}t)-\textrm{Si}(Ae^{-\frac{w}{\lambda}}t))],
	\label{eq:St}
\end{align}
where $\textrm{Ci}(x)$ is Cosine Integral and defined as $\textrm{Ci}(x)=-\int^{\infty}_{x}\frac{\cos t}{t}dt$; $\textrm{Si}(x)$ is Sine Integral and defined as $\textrm{Si}(x)=\int^{x}_{0}\frac{\sin t}{t}dt$ \cite{Abramowitz}. One can use Eq. (\ref{eq:St}) to calculate the $\Delta T$ in Eq. (\ref{eq:dfdT}). To the first order, $\Delta T$ can be expressed as
\begin{equation}
	\Delta T=-\frac{\textrm{Si}(Ae^{-\frac{d}{\lambda}}T)-\textrm{Si}(Ae^{-\frac{w}{\lambda}}T)}{\textrm{Ci}(Ae^{-\frac{d}{\lambda}}T)-\textrm{Ci}(Ae^{-\frac{w}{\lambda}}T)}.
	\label{eq:dTvsT}
\end{equation}
Using the value of $d=250~\rm\mu m$ and $w=5.825~\rm cm$, Eq. (\ref{eq:dTvsT}) is plotted in Fig. \ref{fig:dt-T} (the lower red curve). It is surprising to find that $\Delta T$ does not increase linearly with respect to $T$ all the way up. At certain point, $\Delta T$ becomes more or less a constant, which means the phase difference due to the short-range force stop accumulating after certain time. A longer measurement time will not increase the phase difference due to the short-range force. This behavior is the result of the inhomogeneous broadening due to the varying distance between the $^3$He spin and the mass. 

This simple approximation yields a satisfactory result in terms of the time at which the linear relationship breaks down (The exact solution is plotted as a black curve in Fig. \ref{fig:dt-T}). In our experiment, the maximum measurement time is less than 0.2 s to guarantee that the FID measurement is in the linear region, so that the frequency comparison is valid even the measurement time for each configuration is slightly different. 

\begin{figure}[htbp]
	\centering
		\includegraphics[width=0.48\textwidth]{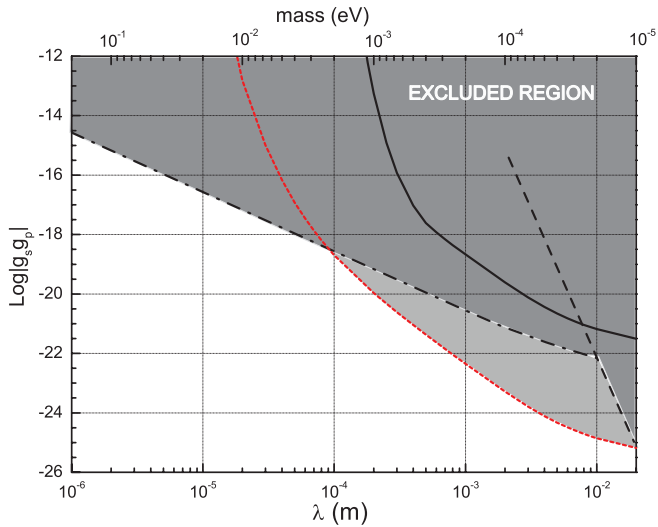}
	\caption{(Color online) Constraints on the coupling constant product $g_sg_p$ of the spin-dependent force as a function of the range $\lambda$ and the equivalent mass of the axion-like particle mediating the short-range interaction. The dashed line is the result from \cite{Youdin}, the dash-dotted line is the re-analysis of the $T_2$ measurements of \cite{Gemmel} by \cite{Petukhov}, the solid line is the analysis of our present experiment, and the dotted line is a projected sensitivity achievable using our method based on the stability of the magnetic field demonstrated in \cite{Chupp}. The dark gray is the excluded region and the light gray is the region that could be excluded with the improved field stability.}
	\label{fig:limits}
\end{figure}

By choosing different values of $\lambda$, the constraints on the coupling constants $g_sg_p$ are found and plotted as a solid line in Fig. \ref{fig:limits}. The 250 $\mu$m window thickness of the double chamber glass cell allows us to constrain interactions ranges $\lambda$ down to $10^{-4}$ m. The dominant source of the uncertainty in our experiment came from magnetic field fluctuations. The gradient compensation coils, needed to achieve an uniform field in our apparatus, added uncorrelated magnetic field noise to the holding field as the gradient coils were powered by independent power supplies.   

There are many avenues for the improvement of the measurement sensitivity using this technique. One can use a dedicated $^3$He cell with a shorter length and add magnetic shielding instead of gradient coils to improve the field uniformity and stability. In \cite{Chupp}, the authors conducted a precision frequency measurement using polarized gases in an apparatus with three layers of cylindrical $\mu$ metal shielding for field uniformity and a co-magnetometer technique to reduce the magnetic field noise by 3 to 4 orders of magnitude. They achieved a precision of $10^{-6}$ Hz out of 1000 Hz, two orders of magnitude better than our pilot experiment. In Fig. \ref{fig:limits} we show the limits on $g_sg_p$ (dotted line) which could be achieved with this technique using the sensitivity demonstrated in \cite{Chupp} assuming all other geometric parameters (window thickness, pickup coil size, etc.) are unchanged. As shown in Fig \ref{fig:limits}, this projected sensitivity would represent a significant improvement in the $\lambda$ region of $10^{-2}$ to $10^{-4}$ m, compared with the best existing laboratory limit derived by the authors of \cite{Petukhov}, shown as the dotted-dashed line, based on the results of a $T_2$ measurement from \cite{Gemmel}.

The sensitivity of the experiment below $\lambda=10^{-4}$ m is clearly limited as the thickness of the window becomes much larger than the force range. The 7 amg $^3$He cell has an internal pressure of more than 10 atm at 200 $\rm ^o$C and the 250 $\mu$m window thickness is needed for strength. However, the high pressure is not necessary for this experiment. A 1 amg $^3$He cell with reduced window thickness could be used to improve the sensitivity of the measurement, especially in the $\lambda<10^{-4}$ m range. Another order of magnitude improvement on the sensitivity could be achieved if denser material was used as the unpolarized mass, such as pure copper or tungsten. We conclude that our proposed method shows a promising sensitivity, with at least one to two orders of magnitude improvement over the current best limit possible in a dedicated experiment with better magnetic field stability. A even higher sensitivity could be achieved if a thinner wall for the $^3$He cell and denser material is used.

The authors want to thank Mike Souza and Gordon Cates' group for their help with the cell construction and Todd Averett's group for the cell filling. This work was supported by the School of Arts and Sciences of Duke University and the U.S. Department of Energy under Contract No. DE-FG02-03ER41231. C. Fu and W. M. Snow acknowledge support from the U. S. National Science Foundation through grant PHY-1068712. M. Snow acknowledges support from the Indiana University Center for Spacetime Symmetries.

\end{document}